\documentstyle[prl,aps,epsf,prbbib]{revtex}

\newcommand{\lapprox}{\lesssim}

\begin{document}
\draft
\twocolumn[\hsize\textwidth\columnwidth\hsize\csname  
@twocolumnfalse\endcsname

\title{From the Bogoliubov-de Gennes equations for
superfluid fermions to the Gross-Pitaevskii equation for condensed bosons}
\author{P. Pieri and G.C. Strinati}
\address{Dipartimento di Fisica, UdR INFM,
 Universit\`{a} di Camerino, I-62032 Camerino, Italy}
\date{\today}
\maketitle
\hspace*{-0.25ex}

\begin{abstract}
We derive the time-independent Gross-Pitaevskii equation at zero
temperature for condensed bosons, which
form as bound-fermion pairs when the mutual fermionic attractive
interaction is sufficiently strong, from
the strong-coupling limit of the Bogoliubov-de Gennes equations that
describe superfluid fermions in
the presence of an external potential.
Three-body corrections to the Gross-Pitaevskii equation are also obtained by
our approach.
Our results are relevant to the recent advances with ultracold fermionic 
atoms in a trap.
\end{abstract}

\hspace*{-0.25ex}
]
\narrowtext

Evolution from superfluid fermions to condensed composite bosons
appears on the verge of being
experimentally realized in terms of dilute ultracold fermionic atoms in a
trap, although several
experimental difficulties still remain to be overcome \cite{Inguscio-02}.
Dilute condensed bosons in harmonic traps, in particular, have already been
studied extensively \cite{dilute-bosons}.
From a theoretical point of view, their macroscopic properties (such as the
density profile) have been described
by the time-independent Gross-Pitaevskii (GP) equation for the condensate
wave function \cite{G-P-original}
(see also Refs.~\onlinecite{DGPS-99} and \onlinecite{Pethick-Smith}).
Superfluid fermions in the presence of a spatially varying external
potential, on the other hand, are
usually described in terms of the Bogoliubov-de Gennes (BdG) equations
\cite{DeGennes}, where a spatially varying
gap function is obtained from a set of two-component fermion wave functions.
Several problems (such as the calculation
of the Josephson current \cite{BTK})
have been approached in terms of the BdG equations for
superfluid fermions.

Given the possible experimental connection between the properties of
superfluid fermions with
a strong mutual attraction and of condensed bosons, a corresponding
theoretical connection between these two
(GP and BdG) approaches seems appropriate at this time.
In this way, one could even explore the interesting intermediate-coupling
(crossover) region, where neither the
fermionic nor the bosonic properties are fully realized.
Such a connection might be also relevant for the study of high-temperature
(cuprate) superconductors, for which
the coupling strength of the superconducting interaction seems to be
intermediate between weak coupling (with
largely overlapping Cooper pairs) and strong coupling (with nonoverlapping
composite bosons).

In this paper, we establish this connection by showing that the
time-independent GP equation at zero temperature can be obtained as the 
strong-coupling limit (to be specified below) of
the BdG equations.
This result thus shows that the BdG equations, originally conceived for 
weak coupling, are also appropriate to describe the strong-coupling limit, 
whereby the fermionic gap function is suitably mapped onto the
condensate wave function of the GP equation.
Our derivation resembles closely Gorkov's microscopic derivation of the
Ginzburg-Landau equations from
BCS theory near the superconducting critical temperature in the
weak-coupling limit \cite{Gorkov}.
Our approach further enables us to obtain high-order corrections to the GP
equation; in particular, the
three-body correction is here explicitly derived.

We begin by considering the BdG equations for superfluid fermions in
the presence of a spatially varying
external potential $V({\mathbf r})$:

\begin{equation}
\left( \begin{array}{cc}
{\mathcal H}({\mathbf r})  &    \Delta({\mathbf r})   \\
\Delta({\mathbf r})^{*}   &  - {\mathcal H}({\mathbf r})  \end{array}
\right)
\left( \begin{array}{c}
u_{n}({\mathbf r}) \\
v_{n}({\mathbf r}) \end{array}
\right)
\, = \, \epsilon_{n} \,
\left( \begin{array}{c}
u_{n}({\mathbf r}) \\
v_{n}({\mathbf r}) \end{array}
\right) \,\, .
\label{BdG-equations}
\end{equation}

\noindent
Here
\begin{equation}
{\mathcal H}({\mathbf r}) \, = \, - \frac{\nabla^{2}}{2m} \, + \,
V({\mathbf r}) \, - \, \mu
\label{single-particle-Hamiltonian}
\end{equation}

\noindent
is the single-particle Hamiltonian reckoned on the (fermionic) chemical
potential $\mu$ ($m$ being the fermionic
mass and $\hbar = 1$ throughout), while

\begin{equation}
\Delta({\mathbf r}) \, = \, - \,  v_{0} \, \, \sum_{n} \, u_{n}({\mathbf r})
\, v_{n}({\mathbf r})^{*}
\left[ 1 \, - \, 2 \, f(\epsilon_{n}) \right]
\label{gap-function}
\end{equation}

\noindent
is the gap function that has to be self-consistently determined
\cite{DeGennes}, where
$f(\epsilon_{n})= \left[ \exp (\beta \epsilon_{n}) + 1 \right]^{-1}$ is the
Fermi function with inverse temperature
$\beta$ (the sum in Eq.~(\ref{gap-function}) being limited to positive
eigenvalues $\epsilon_{n}$ only).
The negative constant $v_{0}$ in Eq.~(\ref{gap-function}) originates from
the attractive interaction
acting between fermions with opposite spins (or fermionic atoms with two 
different internal states) and taken of the contact-potential form
$v_{0} \, \delta({\mathbf r} - {\mathbf r}')$, which can be conveniently
regularized in terms of the scattering length
$a_{F}$ of the associated two-body problem\cite{Randeria-93,Pi-S-98}.
Correspondingly, the gap function has $s$-wave component only with spinless
structure.

Solution of the BdG equations (\ref{BdG-equations}) is equivalent to
considering the associated Green's function
equation (in matrix form):

\begin{equation}
\left( \begin{array}{cc}
i \omega_{s}  - {\mathcal H}({\mathbf r})  & - \Delta({\mathbf r})   \\
- \Delta({\mathbf r})^{*}   &  i \omega_{s} + 
{\mathcal H}({\mathbf r})  \end{array}
\right)
\hat{{\mathcal G}}({\mathbf r},{\mathbf r}';\omega_{s})
 = \hat{{\mathbf 1}} \delta({\mathbf r} - {\mathbf r}')
\label{Green-function-equations}
\end{equation}

\noindent
where $\omega_{s} = (2s+1)\pi/\beta$ ($s$ integer) is a fermionic Matsubara
frequency, $\hat{{\mathbf 1}}$ is
the unit dyadic, and $\hat{{\mathcal G}}$ is the single-particle Green's
function in Nambu's notation \cite{Schrieffer}.
Solutions of Eqs.~(\ref{BdG-equations}) and
(\ref{Green-function-equations}) are, in fact, related by the expression:

\begin{eqnarray}
& &\hat{{\mathcal G}}({\mathbf r},{\mathbf r}';\omega_{s})  =   \sum_{n} \,
\left( \begin{array}{c}
u_{n}({\mathbf r}) \\
v_{n}({\mathbf r}) \end{array}
\right) \, \frac{1}{i \omega_{s} \, - \, \epsilon_{n}} \,
\left( u_{n}({\mathbf r}')^{*} , v_{n}({\mathbf r}')^{*} \right)
\nonumber \\
& &+  \sum_{n} \,
\left( \begin{array}{c}
- v_{n}({\mathbf r})^{*} \\
u_{n}({\mathbf r})^{*} \end{array}
\right) \, \frac{1}{i \omega_{s} \, + \, \epsilon_{n}} \,
\left( - v_{n}({\mathbf r}') , u_{n}({\mathbf r}') \right) \,\, .
\label{Green-function-BdG}
\end{eqnarray}

We adopt at this point Gorkov's procedure \cite{Gorkov} for expressing the
solution of Eq.~(\ref{Green-function-equations})
in terms of the non-interacting Green's function $\tilde{{\mathcal G}_{0}}$ 
that satisfies the equation

\begin{equation}
\left[ i \omega_{s} \, - \, {\mathcal H}({\mathbf r}) \right] \,
\tilde{{\mathcal G}_{0}}({\mathbf r},{\mathbf r}';\omega_{s})
\, =  \, \delta({\mathbf r} - {\mathbf r}')
\label{Green-function-0}
\end{equation}

\noindent
and subject to the \emph{same\/} external potential $V({\mathbf r})$.
We are thus led to consider the two coupled integral equations:

\begin{eqnarray}
{\mathcal G}_{11}({\mathbf r},{\mathbf r}';\omega_{s})  & = &
\tilde{{\mathcal G}_{0}}({\mathbf r},{\mathbf r}';\omega_{s})
\, + \, \int \! d{\mathbf r}'' \,
\tilde{{\mathcal G}_{0}}({\mathbf r},{\mathbf r}'';\omega_{s}) \,
\Delta({\mathbf r}'')\nonumber \\
&\times& {\mathcal G}_{21}({\mathbf r}'',{\mathbf r}';\omega_{s})
\label{coupled-integral-equation-1}   \\
{\mathcal G}_{21}({\mathbf r},{\mathbf r}';\omega_{s})  & = & - \int \!
d{\mathbf r}'' \,
\tilde{{\mathcal G}_{0}}({\mathbf r}'',{\mathbf r};-\omega_{s}) 
\Delta({\mathbf r}'')^{*}\nonumber \\
&\times&  {\mathcal G}_{11}({\mathbf r}'',{\mathbf r}';\omega_{s})
\label{coupled-integral-equation-2}
\end{eqnarray}

\noindent
for the normal (${\mathcal G}_{11}$) and anomalous (${\mathcal G}_{21}$)
single-particle Green's functions.
Equations (\ref{coupled-integral-equation-1}) and 
(\ref{coupled-integral-equation-2}), together with the 
self-consistency equation
\begin{equation}
\Delta^*({\mathbf r})= v_0\frac{1}{\beta} 
\sum_s {\mathcal G}_{21}({\mathbf r},{\mathbf r};\omega_{s}) 
e^{-i \omega_s \eta}
\label{selfc}
\end{equation}
($\eta \to 0^+$), are fully equivalent
to the original BdG equations (\ref{BdG-equations}) and (\ref{gap-function}), 
and hold for any
coupling.

We pass now to specifically consider the \emph{strong-coupling limit\/} of
Eqs.~(\ref{coupled-integral-equation-1})-(\ref{selfc}), and show under what 
circumstances they reduce to the GP equation for spinless composite bosons 
with mass $2m$ and subject to the potential $2 \,
V({\mathbf r})$. In this limit, the fermionic chemical potential 
approaches $-\varepsilon_0/2$, where $\varepsilon_0=(m a_F^2)^{-1}$ is the binding 
energy of the composite boson which represents the largest energy scale of 
the problem. 

In the present context, achieving the strong-coupling limit implies that 
the following conditions on the potential  
\begin{equation}
|V({\mathbf r})| \ll |\mu|, \, a_{F} |\nabla
V({\mathbf r})| \ll |V({\mathbf r})|, \, a_{F}^{2}|\nabla^{2} V({\mathbf r})| 
\ll |V({\mathbf r})|        \label{conditions-V}
\end{equation}

\noindent
hold for all relevant values of ${\mathbf r}$.
In the strong-coupling limit, $a_{F} \sim (2m|\mu|)^{-1/2}$ represents the 
characteristic length scale
for the non-interacting Green's function of the associated homogeneous
problem (with $V({\mathbf r})=0$),
as it can be seen from the expression

\begin{eqnarray}
& &{\mathcal G}_{0}({\mathbf r} - {\mathbf r}';\omega_{s}|\mu) = \int \!
\frac{d{\mathbf k}}{(2 \pi)^{3}} 
\frac{e^{i{\mathbf k}\cdot({\mathbf r}-{\mathbf r}')}}{i \omega_{s}  - 
{\mathbf k}^{2}/(2m)  + \mu} \nonumber \\
& & =  - \frac{m}{2\pi |{\mathbf r}-{\mathbf r}'|}  \exp \left\{ i \,
{\mathrm sgn}(\omega_{s})
\sqrt{2m(\mu + i \omega_{s})}|{\mathbf r}-{\mathbf r}'| \right \}
\label{non-interacting-Green-function}
\end{eqnarray}

\noindent
that holds for any coupling.
With the conditions (\ref{conditions-V}), it can be readily verified that
the position

\begin{equation}
\tilde{{\mathcal G}_{0}}({\mathbf r},{\mathbf r}';\omega_{s}) \simeq 
{\mathcal G}_{0}({\mathbf r} - {\mathbf r}';\omega_{s}|\mu -
(V({\mathbf r})+V({\mathbf r}'))/2),
\label{approximate-G-0}
\end{equation}

\noindent
whereby the chemical potential $\mu$ in expression
(\ref{non-interacting-Green-function}) is replaced by
the local form $\mu - (V({\mathbf r})+V({\mathbf r}'))/2$, satisfies
Eq.~(\ref{Green-function-0}) in the presence of
the external potential.
[The midpoint rule, albeit superfluous for most of the following arguments
owing to the slowness of the potential over
the distance $a_{F}$, has been adopted for later convenience in the
derivation of the current density.]
The approximate expression (\ref{approximate-G-0}) plays in the present
context an analogous role to the eikonal
approximation for Gorkov's problem in the presence of an external magnetic
field \cite{Gorkov}.

From a physical point of view, the first of the conditions
(\ref{conditions-V}) implies that the external potential
is much smaller than the binding energy $\varepsilon_0$, so
that the composite boson does not break 
apart by scattering against the potential $V({\mathbf r})$.
The two additional conditions (\ref{conditions-V}), on the other hand,
imply that the external potential is
sufficiently slowly varying over the characteristic size $a_{F}$ of the
composite boson, so that the composite boson
itself probes the potential as if it were a point-like particle.

In particular, for the corresponding one-dimensional problem with a generic
(albeit non pathological) potential $V(x)$
that satisfies the first two conditions (\ref{conditions-V}) ($a_{F}$
therein being replaced by $a=(2m|\mu|)^{-1/2}$),
one can readily verify that our position (\ref{approximate-G-0}), namely,

\begin{equation}
\tilde{{\mathcal G}_{0}}(x,x';\omega_{s}) \, = \, - \,
\frac{m \, e^{ - \sqrt{2m(|\mu|+(V(x)+V(x'))/2 -i\omega_{s})} \, |x-x'|}}{\sqrt{2m(|\mu|
+(V(x)+V(x'))/2-i\omega_{s})}} \,
\label{1D-ours}
\end{equation}
\noindent
is fully equivalent to the solution of the (one-dimensional version of the)
Green's function equation
(\ref{Green-function-0}) as obtained in terms of the asymptotic ($a
\rightarrow 0^{+}$) WKB approximation
\cite{Bender-Orszag}.
This identification holds in the relevant range $|x-x'| \lapprox a$ and
provided $|\omega_{s}|\ll|\mu|$
(that can be always satisfied for sufficiently large $|\mu|$).

Derivation of the GP equation from the BdG equations in the
strong-coupling limit exploits the position (\ref{approximate-G-0}) and 
proceeds by using Eq.~(\ref{coupled-integral-equation-1}) to expand
perturbatively Eq.~(\ref{coupled-integral-equation-2}) up to third order in
$\Delta({\mathbf r})$ (the ratio $\Delta({\mathbf r})/|\mu|$ provides 
the small parameter for this expansion, by
taking advantage of  the ``diluteness'' condition of the system).
Equation (\ref{selfc}) yields eventually the following integral equation for 
the gap function:

\begin{eqnarray}
& &- \, \frac{1}{v_{0}} \, \Delta({\mathbf r})^{*} =   \int \!
d{\mathbf r}_{1} \,\,
Q({\mathbf r},{\mathbf r}_{1}) \,\, \Delta({\mathbf r}_{1})^{*}
\nonumber \\
& &+  \int \! d{\mathbf r}_{1} d{\mathbf r}_{2} d{\mathbf r}_{3} \,\,
R({\mathbf r},{\mathbf r}_{1},{\mathbf r}_{2}.{\mathbf r}_{3}) \,\,
\Delta({\mathbf r}_{1})^{*} \, \Delta({\mathbf r}_{2}) \,
\Delta({\mathbf r}_{3})^{*}           
\label{integral-eq-Delta}
\end{eqnarray}

\noindent
with the notation

\begin{equation}
Q({\mathbf r},{\mathbf r}_{1})  \, = \, \frac{1}{\beta} \sum_{s} \,
\tilde{{\mathcal G}_{0}}({\mathbf r}_{1},{\mathbf r};-\omega_{s}) \,
\tilde{{\mathcal G}_{0}}({\mathbf r}_{1},{\mathbf r};\omega_{s})
\label{definition-Q}
\end{equation}

\noindent
and

\begin{eqnarray}
R({\mathbf r},{\mathbf r}_{1},{\mathbf r}_{2},{\mathbf r}_{3}) & = & - \,
\frac{1}{\beta} \sum_{s} \,
\tilde{{\mathcal G}_{0}}({\mathbf r}_{1},{\mathbf r};-\omega_{s}) \,
\tilde{{\mathcal G}_{0}}({\mathbf r}_{1},{\mathbf r}_{2};\omega_{s})\nonumber\\
&\times& \tilde{{\mathcal G}_{0}}({\mathbf r}_{3},{\mathbf r}_{2};-\omega_{s}) \,
\tilde{{\mathcal G}_{0}}({\mathbf r}_{3},{\mathbf r};\omega_{s}) \,\, .
\label{definition-R}
\end{eqnarray}

\noindent
The position (\ref{approximate-G-0}) implies that, in strong coupling, 
$Q({\mathbf r},{\mathbf r}_1)$ vanishes for
$|{\mathbf r}-{\mathbf r}_1| \gtrsim a_F$. 
Since (as a consequence of the conditions (\ref{conditions-V})) 
$\Delta({\mathbf r})$ is slowly varying over the length scale $a_{F}$, we can 
set $\Delta({\mathbf r}_1)^* \simeq \Delta({\mathbf r})^*$ on the left-hand 
side of Eq.~(\ref{integral-eq-Delta}) and write
\begin{equation}
\int \! d{\mathbf r}_{1} Q({\mathbf r},{\mathbf r}_{1}) 
\Delta({\mathbf r}_1)^{*} \simeq 
a_{0}({\mathbf r}) \Delta({\mathbf r})^{*} + 
\frac{1}{2} b_{0}({\mathbf r}) \nabla^{2} \Delta({\mathbf r})^{*}
\label{approximate-Q-Delta}
\end{equation}

\noindent
where

\begin{eqnarray}
a_{0}({\mathbf r}) &=&  \int \! d{\mathbf r}_1 
Q({\mathbf r},{\mathbf r}_1)=\int \! 
\frac{d{\mathbf k}}{(2\pi)^3}\frac{1}{\frac{k^2}{m}+2|\mu({\mathbf r})|}
\nonumber\\
  &\simeq& 
-  \, \frac{1}{v_{0}}  +  \frac{m^{2} a_{F}}{8  \pi}  \left(
\mu_{B}  - 2 V({\mathbf r}) \right)
\label{a-0}
\end{eqnarray}

\noindent
and
\begin{equation}
b_{0}({\mathbf r}) \, = \, \frac{1}{3} \, \int \! d{\mathbf r}_1 \,\,
Q({\mathbf r},{\mathbf r}_1) \, |{\mathbf r}_1-{\mathbf r}|^{2}
\, \simeq \, \frac{m \, a_{F}}{16 \, \pi} \,\, .
\label{b-0}
\end{equation}

\noindent
Here, $\mu({\mathbf r})=\mu -V({\mathbf r})$ and we have introduced the 
bosonic chemical potential $\mu_{B} = 2 \mu + \varepsilon_0$ such 
that $\mu_{B}\ll\varepsilon_0$. Note that $(V({\mathbf r})+V({\mathbf r}_1))/2$
in the local chemical potential has been
replaced by $V({\mathbf r})$ at the relevant order, since the external
potential is slowly varying over the distance $a_{F}$.
Disposal of the ultraviolet divergence in Eq.~(\ref{a-0}) via the 
regularization of the contact potential\cite{Randeria-93,Pi-S-98} in terms 
of $a_{F}$ and use 
of the strong-coupling assumption $\beta\mu\to -\infty$ have led to 
the approximate expressions (\ref{a-0}) and (\ref{b-0}).
By a similar token, we obtain

\begin{eqnarray}
& &\int \! d{\mathbf r}_{1} d{\mathbf r}_{2} d{\mathbf r}_{3} \,\,
R({\mathbf r},{\mathbf r}_{1},{\mathbf r}_{2},{\mathbf r}_{3}) \,\,
\Delta({\mathbf r}_{1})^{*} \, \Delta({\mathbf r}_{2}) \,
\Delta({\mathbf r}_{3})^{*}\nonumber \\
& &\, \simeq \,
c_{0} \, |\Delta({\mathbf r})|^{2} \, \Delta({\mathbf r})^{*}
\label{approximate-R-Delta}
\end{eqnarray}

\noindent
with

\begin{equation}
c_{0} \, \simeq \, - \, \left( \frac{m^{2} \, a_{F}}{8 \, \pi}
\right)^{2} \,
\frac{8 \, \pi \, a_{F}}{2 \, m}  \,\, .
\label{c-0}
\end{equation}

Entering the approximate expressions
(\ref{approximate-Q-Delta})-(\ref{c-0}) into the integral equation
(\ref{integral-eq-Delta}) for the gap function and introducing the
\emph{condensate wave function\/}

\begin{equation}
\Phi({\mathbf r}) \, = \, \sqrt{\frac{m^{2} \, a_{F}}{8 \, \pi}} \,
\Delta({\mathbf r})  \,\, ,   \label{wave-function}
\end{equation}

\noindent
we obtain eventually:
\begin{equation}
- \frac{1}{4 m}  \nabla^{2} \Phi({\mathbf r})  +  2 
V({\mathbf r})  \Phi({\mathbf r}) + 
\frac{8 \pi a_{F}}{ 2 m}  |\Phi({\mathbf r})|^{2} 
\Phi({\mathbf r}) = 
\mu_{B} \Phi({\mathbf r})  .
\label{G-P-equation}
\end{equation}

\noindent
This is just the {\em time-independent Gross-Pitaevskii equation} for 
composite bosons of mass $m_{B}=2m$, chemical
potential $\mu_{B}$, subject to the external potential $2
\, V({\mathbf r})$, and mutually interacting
via the short-range repulsive potential $4 \pi a_{B}/m_{B}$ 
(where $a_{B}=2a_{F}$ is the bosonic scattering length within the Born 
approximation \cite{Haussmann,Pi-S-98}).
Note that the two-body binding
energy $\varepsilon_0$ has been eliminated from
explicit consideration via the bound-state equation.
Note also that the nontrivial rescaling in Eq.~(\ref{wave-function}) between 
the gap function and the condensate wave function has been fixed by the 
nonlinear term in Eq.~(\ref{G-P-equation}).

Equation~(\ref{G-P-equation}) has been formally obtained from the original 
BdG equations in the limit $\beta\mu\to - \infty$. In this respect, it  
would seem that this equation holds even near the condensation temperature 
$T_c$, where the GP equation is instead known not to be valid.
On physical grounds, however, at finite temperature excitations of bosons out
of the condensate (that are not included in the present treatment) are 
expected to be
important especially in the strong-coupling regime of the BdG equations, 
thus restricting the range of validity of Eq.~(\ref{G-P-equation}) near zero 
temperature.

The physical interpretation of the condensate wave function
(\ref{wave-function}) can be obtained from the
general expression for the density
$n({\mathbf r}) = (2/\beta) \sum_{s} \exp (i \omega_{s} \eta )
{\mathcal G}_{11}({\mathbf r},{\mathbf r};\omega_{s})$,
where ${\mathcal G}_{11}$ is obtained by combining
Eqs.~(\ref{coupled-integral-equation-1}) and
(\ref{coupled-integral-equation-2}) and expanding perturbatively up to
second order in $\Delta({\mathbf r})$.
Recalling that the fermionic contribution
$(2/\beta) \sum_{s} \exp (i \omega_{s} \eta )
{\mathcal G}_{0}({\mathbf r},{\mathbf r};\omega_{s})$ vanishes in the
strong-coupling limit, one is left with the expression

\begin{eqnarray}
n({\mathbf r}) & \simeq & - \, 2 \, |\Delta({\mathbf r})|^{2} \, \int \!
\frac{d{\mathbf k}}{(2\pi)^{3}} \, \frac{1}{\beta}
\sum_{s} \, {\mathcal G}_{0}({\mathbf k},\omega_{s})^{2} \,
{\mathcal G}_{0}({\mathbf k},-\omega_{s})        \nonumber \\
& \simeq & - \, 2 \, |\Delta({\mathbf r})|^{2} \, \left( - \, \frac{m^{2} \,
a_{F}}{8 \, \pi} \right)
\, = 2 \, |\Phi({\mathbf r})|^{2}
\label{n-wave-function}
\end{eqnarray}

\noindent
evaluated in terms of the wave-vector representation of ${\mathcal G}_{0}$.
Since the bosonic density $n_{B}({\mathbf r})$ is just half the fermionic
density $n({\mathbf r})$, from
Eq.~(\ref{n-wave-function}) we obtain that $n_{B}({\mathbf r}) =
|\Phi({\mathbf r})|^{2}$, as usual with the GP equation.
The normalization condition $N = \int \! d{\mathbf r} \, n({\mathbf r})$ (or,
equivalently, its bosonic counterpart) fixes,
in turn, the overall chemical potential $\mu$ (or directly $\mu_{B}$ in the
strong-coupling limit).

By a similar token, the current density can be obtained from the general
expression

\begin{equation}
{\mathbf j}({\mathbf r})  = \frac{1}{i m} \left( \nabla  - 
\nabla' \right) 
\frac{1}{\beta} \sum_{s} e^{i \omega_{s} \eta}  
{\mathcal G}_{11}({\mathbf r},{\mathbf r}';\omega_{s})
|_{{\mathbf r}={\mathbf r}'}
\label{j-wave-function}
\end{equation}

\noindent
by combining Eqs.~(\ref{coupled-integral-equation-1}) and
(\ref{coupled-integral-equation-2}) as before.
The independent-particle contribution to the current now vanishes exactly
with the midpoint rule
(\ref{approximate-G-0}).
After long but straightforward manipulations one obtains for the remaining 
contributions in the limit $\beta \mu \to - \infty$:

\begin{equation}
{\mathbf j}({\mathbf r}) \, \simeq \, \frac{1}{2 i m} \, \left[
\Phi({\mathbf r})^{*} \, \nabla \, \Phi({\mathbf r}) \, - \,
\Phi({\mathbf r}) \, \nabla \, \Phi({\mathbf r})^{*} \right]
\label{j-GP}
\end{equation}

\noindent
which is, as expected, twice the value of the quantum-mechanical expression
of the current for a composite
boson with mass $m_{B}=2m$ and wave function $\Phi({\mathbf r})$.

It is clear from the above derivation that higher-order corrections to the
GP equation (\ref{G-P-equation})
can also be obtained by expanding Eqs.~(\ref{coupled-integral-equation-1})
and (\ref{coupled-integral-equation-2})
to higher than the third order in $\Delta({\mathbf r})$.
In particular, to fifth order in $\Delta({\mathbf r})$ the following term
adds to the right-hand side of 
Eq.~(\ref{coupled-integral-equation-2}) (with ${\mathbf r}={\mathbf r}'$)
once summed over $\omega_s$:

\begin{equation}
-  |\Delta({\mathbf r})|^{4} \Delta({\mathbf r})^{*} \int \!
\frac{d{\mathbf k}}{(2\pi)^{3}} \frac{1}{\beta}
\sum_{s}  {\mathcal G}_{0}({\mathbf k},\omega_{s})^{3} 
{\mathcal G}_{0}({\mathbf k},-\omega_{s})^{3} .
\label{5-order-Delta}
\end{equation}

\noindent
Evaluating the integral in Eq.~(\ref{5-order-Delta}) in the strong-coupling
limit and recalling the rescaling
(\ref{wave-function}), one finds that the term $g_{3}
|\Phi({\mathbf r})|^{4} \Phi({\mathbf r})$ adds to
the left-hand side of the GP equation (\ref{G-P-equation}), where the
\emph{three-body interaction\/}
$g_{3}= - 30 \pi^{2} a_{F}^{4}/m$ is attractive.
As an example, taking for $m_{B}$ the mass of $^{85}Rb$ and the typical
value $a_{B} \sim 250 \, a.u.$, one gets
$|g_{3}|/\hbar \sim 10^{-27} cm^{6} s^{-1}$ with the correct order of
magnitude \cite{Kohler-Greene}.

Even further higher-order corrections can be evaluated in this way with not
much additional burden.
Alternatively, when corrections to the GP equation for composite bosons are
needed, one could directly proceed by
evaluating numerically the BdG equations (\ref{BdG-equations}) in the
intermediate- to strong-coupling region,
thus calculating the physical quantities of interest in terms of the
original fermionic wave functions.
Such an approach would be especially useful for addressing the intermediate
(crossover) region, where neither the
fermionic nor the bosonic character of the system are fully developed.

Specifically, the results of the present paper could be applied to 
determine the evolution of the density profile for a system of superfluid 
fermionic atoms in a trap when the effective fermionic attraction is increased.
It has recently been shown\cite{Perali} 
that, as far as the gross features of the density profile are concerned, this 
problem can 
be dealt with by solving the coupled mean-field BCS equations for the gap and 
the density within a local-density approximation. 
Such an approach reduces to 
the (bosonic) Thomas-Fermi approximation in the strong-coupling limit, thus 
missing the contribution of  the kinetic energy. 
The present derivation of the GP equation from the BdG equations 
shows, in this respect, that solution of the BdG equations for fermionic 
atoms in a trap represents a refined approach that correctly treats the 
kinetic energy for all couplings.   

The connection we have established between the BdG and the GP equations
might be also useful for determining the limiting strong-coupling 
behavior of superfluid fermions in the presence of an external 
potential, as, for instance, when studying strong-coupling effects in the 
Josephson and related problems.

In conclusion, we have shown that the time-independent GP equation at zero 
temperature for composite bosons (formed as bound-fermion
pairs) can be obtained on general grounds from the BdG equations
describing superfluid fermions subject to an external potential, in the 
strong-coupling limit of the mutual fermionic attraction.
Corrections to the GP equation have also been derived in powers of the
condensate wave function.

We are indebted to D. Neilson for discussions.


\end{document}